\documentclass[journal=mamobx,manuscript=article]{achemso}
\usepackage[version=3]{mhchem} 
\usepackage{graphicx}
\usepackage{caption}
\usepackage{subcaption}
\usepackage{natbib}
\usepackage{flexisym}

\author{Bekele J. Gurmessa}
\affiliation[USD]{Department of Physics and Biophysics, University of San Diego, San Diego, United States}
\author{Nicholas Bitten} 
\affiliation{School of Physics and Astronomy, Rochester Institute of Technology, Rochester}
\author{Dan T. Nguyen}
\affiliation{Department of Materials \& BMSE Program, University of California,  Santa Barbara}
\author{Jennifer L. Ross}
\affiliation{Department of Physics, University of Massachusetts, Amherst}
\author{Omar A. Saleh} 
\affiliation{Department of Materials \& BMSE Program, University of California,  Santa Barbara}
\author{Moumita Das}
\affiliation{School of Physics and Astronomy, Rochester Institute of Technology, Rochester}
\author{Rae M. Robertson-Anderson}
\affiliation[USD]{Department of Physics and Biophysics, University of San Diego, San Diego, United States}
\email {randerson@sandiego.edu}
\title{Active Disassembly and Reassembly of Actin Networks Induces Distinct Biphasic Mechanics}

\begin{document}
\begin{abstract}
Actin is a key component of the cytoskeleton, which plays central roles in cell motility, division, growth, and tensile strength. To enable this wide range of transient mechanical processes and properties, networks of actin filaments continuously disassemble and reassemble via active de/re-polymerization. However, the question remains as to how de/re-polymerization kinetics of individual actin filaments translate to time-varying mechanics of dis/re-assembling networks. To address this question, to ultimately elucidate the molecular mechanisms that enable cells to exhibit a myriad of transitory mechanical properties, we couple time-resolved active microrheology with microfluidics to measure the time-varying viscoelastic moduli of entangled and crosslinked actin networks during chemically-triggered network disassembly and reassembly. We also develop a corresponding mathematical model that relates the time-evolution of filament length to the resulting time-dependent storage moduli of evolving networks. During disassembly, we find that the moduli exhibit two distinct exponential decays, with experimental time constants of $\sim$169 and $\sim$47 min, which we show arises from a phase transition from a rigid percolated network to a non-rigid regime occurring after $\sim$90 min. During reassembly, measured moduli exhibit linear increase with time for $\sim$90 min, after which steady-state values are achieved. Our theoretical model shows that reassembly mechanics are dominated by elongation kinetics, and that elongation rates are much slower than has been recently assumed. Our measurements shed much needed light onto how polymerization kinetics map to time-varying mechanics of actively evolving cytoskeleton networks, and provide a powerful platform for studying other active and dynamic systems currently under intense investigation.
\end{abstract}

\section{Introduction}
Actin, a ubiquitous protein comprising the cytoskeleton of cells, plays a key role in orchestrating multifunctional cellular mechanics~\cite{paavilainen2004regulation}. To actively regulate the mechanical properties of the cytoskeleton, actin networks continuously disassemble, reassemble, and reorganize by transitioning between filamentous and monomeric states, thereby enabling vital processes such as mitosis, crawling, and apoptosis~{\citep{stricker2010mechanics, wen2011polymer, gardel2008mechanical, cooper1997cell}. Determining how the mechanics of actin networks evolve during disassembly and reassembly is thus essential for understanding the myriad of time-varying mechanical properties and processes that cells exhibit. Nonetheless, previous studies have focused on either measuring the mechanical properties of steady-state actin networks or the polymerization kinetics of single actin filaments, with the connection between the two still elusive. In particular, how de/re-polymerization kinetics of single actin filaments map to time-varying mechanical properties of actively evolving networks remains an open question.

In the presence of ATP and magnesium, globular actin monomers (G-actin), $\sim$6 $nm$ in size~\citep{byrne2014molecules}, polymerize into semiflexible filaments (F-actin) $\sim$1 - 50 $\mu$m in length with persistence lengths of $\sim$17 $\mu$m~\cite{lieleg2008transient, luan2008micro}. Polymerization proceeds via slow nucleation (i.e. trimer formation), followed by fast elongation, in which ATP-bound monomers bind predominantly to the barbed end of the growing polar filament. Upon depletion of ATP and/or introduction of calcium, F-actin depolymerizes via ATP hydrolysis, release of ADP and P$_i$, and subsequent detachment of G-actin at the pointed end of the filament~\cite{tellam1985mechanism, blanchoin2002hydrolysis}. A number of \emph{in vitro} studies have measured the polymerization and dissociation kinetics of single actin filaments, reporting a wide range of rates and associated mechanisms~\cite{paavilainen2004regulation,fujiwara2002microscopic, pollard1986rate, li2009actin, amann2001direct, blanchoin2002hydrolysis,melki1996continuous,carlier1986direct, fujiwara2007polymerization}. Reported association rate constants span $\sim$1 - 12 $\mu$M$^{-1}$s$^{-1}$ for ATP-bound G-actin on the barbed end, with corresponding elongation/polymerization rates of $\sim$0.03 - 100 $s^{-1}$~\cite{ pollard1986rate, jegou2011individual,crevenna2015side, fujiwara2007polymerization,fujiwara2002microscopic}. Further, light scattering and fluorescence measurements have shown that actin filaments reach $\sim$90\% of their full length in 60 min~\cite{janmey1994mechanical, janmey1988viscoelasticity}. The depolymerization rate for F-actin has been suggested to be rate-limited by the slow release of P$_i$ from actin monomers leading to a rate of 0.16 $s^{-1}$~\cite{fujiwara2007polymerization,jegou2011individual}. However, previous single-molecule experiments found that the length of single depolymerizing actin filaments decreased linearly with time but exhibited multiple linear phases with depolymerization  rates of 0.1 $s^{-1}$ to 1.8 $s^{-1}$~\cite{li2009actin,kueh2008dynamic}. At the same time, similar measurements reported that depolymerization rates increased exponentially with time as filament length decreased~\cite{jegou2011individual}.

High concentrations of F-actin form entangled networks which can also be crosslinked in the presence of actin-binding proteins, leading to unique viscoelastic mechanical properties.  Motivated, in part, by the relevance to cellular mechanics, the viscoelastic response of steady-state entangled and crosslinked actin networks have been extensively studied~\cite{janmey1994mechanical, zaner1995physics, xu1998rheology, ziemann1994local, frey1999viscoelasticity, brau2007passive, lee2010passive}. The majority of studies have focused on characterizing the frequency-dependent storage and loss moduli ($G'$ and $G''$), which quantify the relative elasticity ($G'$), and fluidity ($G''$) of the system~\cite{ziemann1994local,morse1998viscoelasticity,schmidt2000microrheometry,kavanagh1998rheological,gardel2003microrheology}. However, relatively little is known regarding how these properties evolve with time as networks disassemble and reassemble~\cite{janmey1994mechanical,deshpande2015real}. 
To elucidate the underlying dynamic processes the cytoskeleton uses to actively alter its mechanical properties, it is critical to understand how de/re-polymerization kinetics translate to dis/re-assembly of actin networks, and how network dis/re-assembly dynamics map to time-varying mechanical responses of actin networks. 

In this article, we couple time-resolved active microrheology with a recently developed microfluidic platform to measure the time trajectories of local viscoelastic moduli of entangled and crosslinked actin networks during dynamic disassembly and subsequent reassembly. Specifically, we chemically trigger actin de/re-polymerization while simultaneously measuring the force induced in the network by microscale oscillations of an optically trapped microsphere over the time course of dis/re-assembly (Fig.~\ref{ffigschem}). To interpret our results, we develop a mathematical model that integrates polymerization kinetics of actin filaments and the mechanical rigidity of actin networks to describe how the storage modulus of the network varies in time during de/re-polymerization. Polymerization kinetics are modeled using a Master Equation approach \cite{Oosawa,sept1999annealing,ME1,ME2,ME3}, and the mechanics of the network are characterized using an Effective Medium Theory of rigidity percolation \cite{Das1,Das2,Das3,Broedersz}. Our results address key questions regarding the mechanics and corresponding kinetics of dynamically disassembling and reassembling actin networks, elucidating the time-varying mechanics of the cytoskeleton as well as the underlying principles of other active, self-assembling systems of current interest.  

\begin{figure}[htbp!]
\centerline{\includegraphics[width=.5\textwidth]{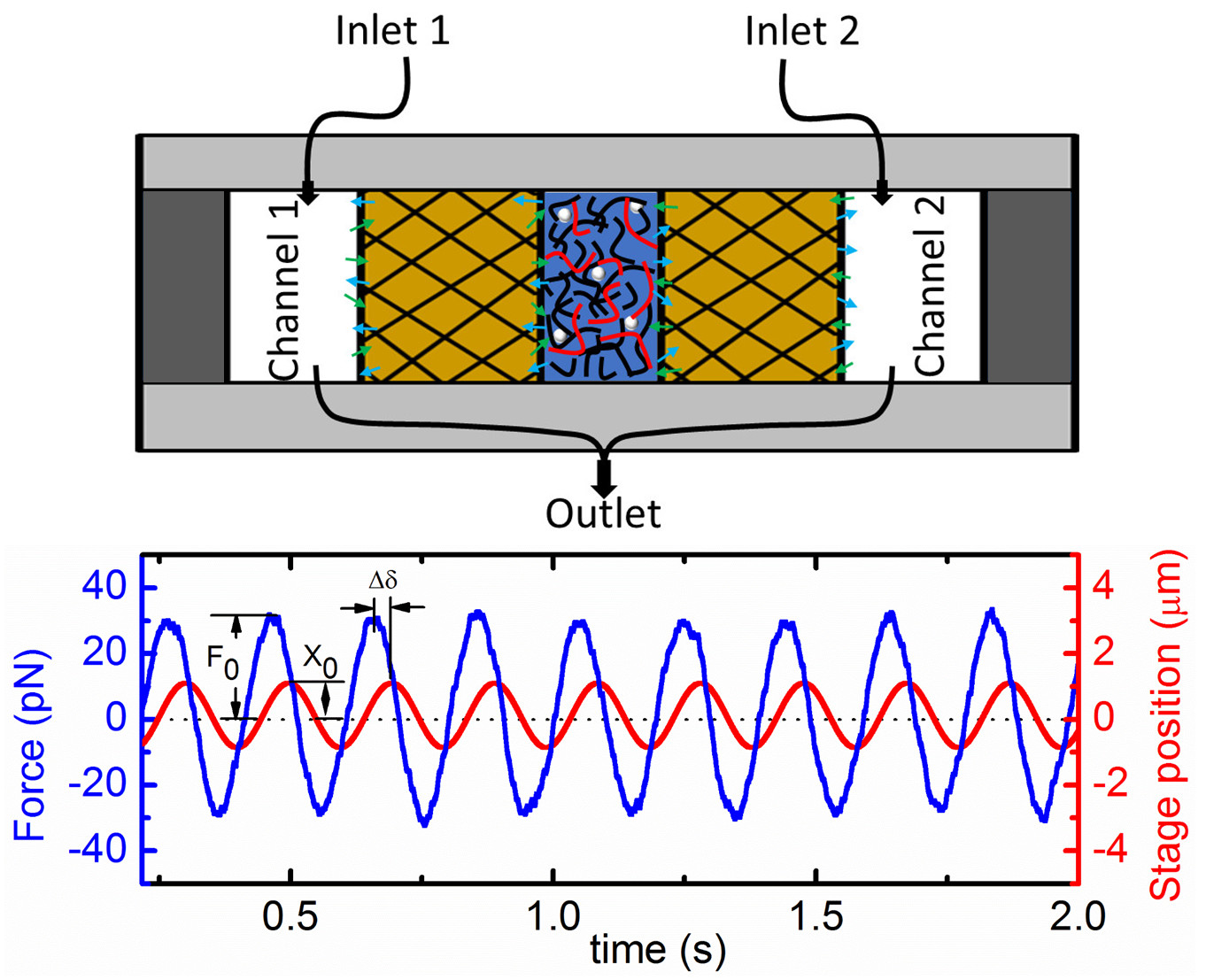}}
\caption{Schematic of experimental approach. (A) Illustration of microfluidic  device comprised of a central channel that contains the actin sample (labeled (red) and unlabeled (black) F-actin and microsphere probes (white)), two flanking channels (1 and 2, white) used for buffer exchange (through inlets 1 \& 2 and an outlet), and two semipermeable membranes (brown) that separate the channels. (B) Sample microrheology data showing probe position  (red) and induced force (blue) as a function of time. The amplitude of the force $F_0$ and position $X_0$ as well as the phase shift $\Delta\delta$ between the force and position are used to determine viscoelastic moduli $G'$ and $G''$.}\label{ffigschem}
\end{figure}

\section*{Materials and Methods}

\textbf{Sample Preparation:} F-actin solutions and crosslinked networks were prepared as previously described~\cite{gurmessa2017nonlinear}. Entangled F-actin solutions were assembled in the sample chamber by polymerizing 0.5 mg/ml unlabeled G-actin (Cytoskeleton) in F-buffer [10 mM Imidazole pH 7.0, 50 mM KCl, 1 mM $\mathrm{MgCl_2}$, 1 mM EGTA, 0.2 mM ATP] for 1 hour at room temperature. 5 $\mu$M pre-assembled Alexa-568-labeled actin filaments and a trace amount of 488-BSA-coated microspheres (4.5 $\mu$m diameter) were also added for visualization and microrheology measurements, respectively. Crosslinked networks were prepared similarly but 0.7 $\mu$M of G-actin was biotinylated and pre-assembled biotin-actin-NeutrAvidin complexes were added at a molar ratio of 0.07 to actin monomers~\citep{gurmessa2017nonlinear}. The length distribution of actin filaments in both networks is 7.2 $\pm$ 4.0 $\mu$m (Fig.~S4), the mesh size is 0.42 $\mu$m (0.3/$\sqrt{c_a}$, with $c_a$ being the actin concentration in mg/ml~\cite{isambert1996dynamics}), and the length between crosslinks for crosslinked networks is $\sim$0.89 $\mu$m~\cite{gurmessa2017nonlinear}.


\textbf{Microfluidics:} The microfluidic device is fabricated as presented in Ref.~\cite{park2016thin} and described further in SI. As shown in Fig.~\ref{ffigschem}, the resulting device is comprised of a central channel, for holding the sample, separated by semipermeable membranes from two flanking buffer channels to enable buffer exchange via diffusion. For measurements, the actin solution was pipetted into the central channel and the flanking channels were initially filled with F-buffer. To initiate disassembly, existing F-buffer is pulled from the flanking channels at a flow rate of 5 $\mu$l/min as G-buffer [2 mM Tris pH 8.0, 0.2 mM ATP, 0.5 mM DTT, 0.1 mM $\mathrm{CaCl_2]}$ is pulled in, thereby enabling diffusion-controlled buffer exchange into the central channel. Complete buffer exchange is achieved in $\sim$6 - 7 min. To trigger reassembly, G-buffer in the flanking channels is exchanged with F-buffer.  

\textbf{Microscopy:}  To independently verify that disassembly occurs over the timescales of our measurements, we use an A1R (Nikon, Melville, NY) laser scanning confocal microscope with 60$x$ 1.4 NA objective to image labeled actin filaments during network disassembly. Specifically, every 20 min during disassembly, time-series are collected by capturing images in 5 s intervals over the course of 3 min. Fig.~\ref{ffigconfoc}, shows the collapse of each time-series. The reduced contrast and increased Brownian noise of images as time proceeds shows that filament mobility increases and network connectivity decreases, demonstrating destruction of a fully-percolated network.

\begin{figure}[htbp!]
\centerline{\includegraphics[height=0.5\textwidth]{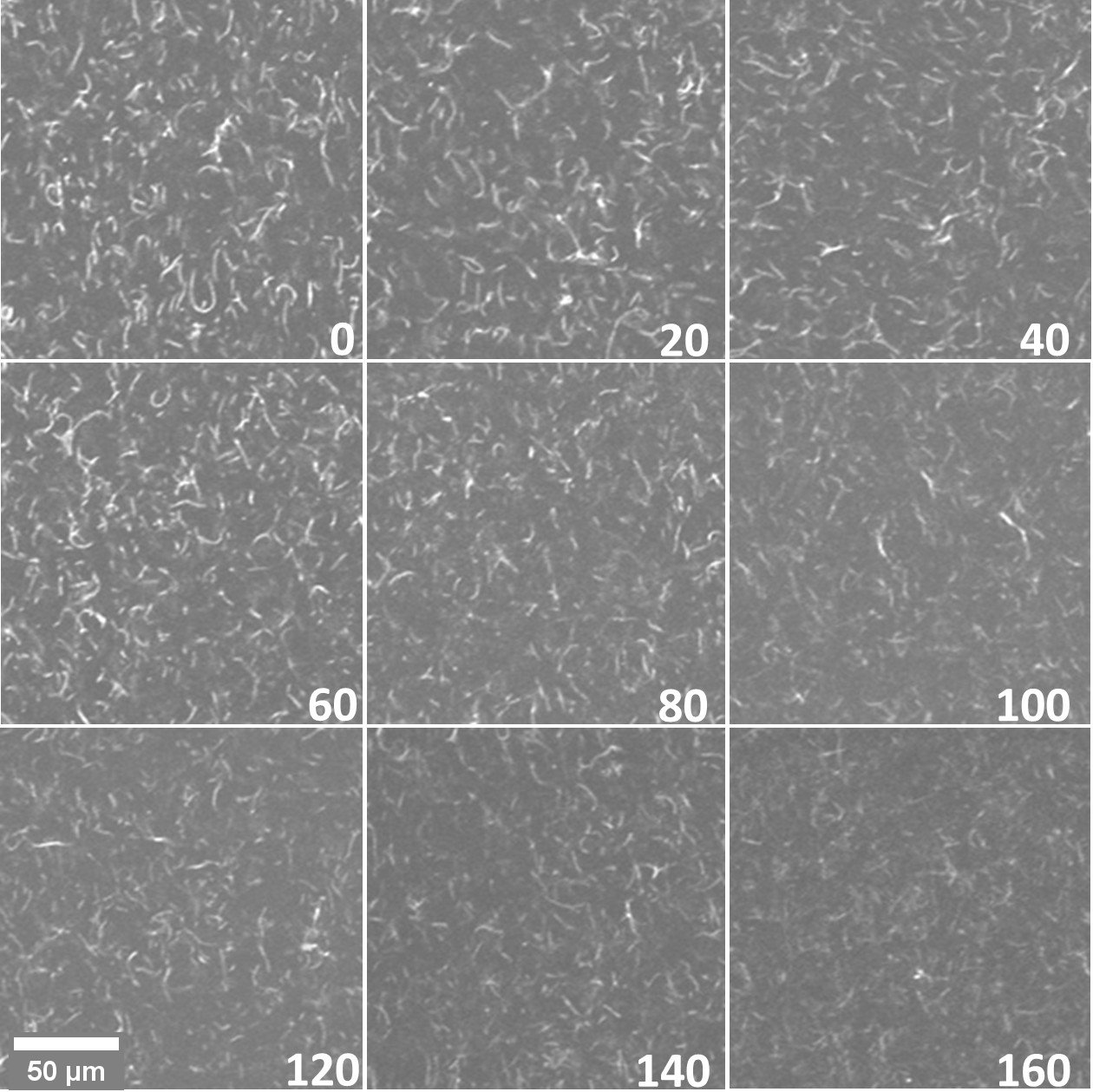}}
\caption{Confocal micrographs of a crosslinked F-actin network during microfluidic-induced depolymerization. Each image is an average of a 3 min time--series of images collected every 5 s using an A1R (Nikon, Melville, NY) laser scanning confocal microscope with 60$x$ 1.4 NA objective. The lower right of each image shows elapsed time in minutes from beginning of network disassembly (i.e. exchange from F-buffer to G-buffer).}\label{ffigconfoc}
\end{figure}

\textbf{Microrheology:} We use an optical trap setup built by outfitting an Olympus IX71 fluorescence microscope with a 1064 nm ND:YAG fiber laser (Manlight) focused with a 60$x$ 1.4 NA objective. A microsphere (probe) embedded in the sample is trapped and oscillated sinusoidally relative to the sample at an amplitude of 1 $\mu$m at five frequencies from 0.5 to 107 rad/s using a piezoelectric nanopositioning stage (Mad City Laboratories) (Fig.~\ref{ffigschem}). A position-sensing detector (Pacific Silicon Sensors) is used to measure the laser deflection, which is proportional to the force on the probe from the surrounding network. The trap stiffness was calibrated via Stokes drag in water~\cite{williams2002optical} and passive equipartition methods~\cite{brau2007passive}. We fit both the stage position and force data to sine curves using the least squares method, and calculate the storage modulus $G'$ and loss modulus $G''$ via $G' = |F_o|\cos(\Delta\delta)/6\pi a |X_o|$ and $G'' = |F_o|\sin(\Delta\delta)/6\pi a |X_o|$ where $F_o$, $X_o$, $\Delta\delta$, $\omega$, and $a$ are the force amplitude, stage position amplitude, phase shift between the force and stage position, oscillation frequency, and probe radius, respectively.  We carried out measurements every 5 min over the time course of disassembly and reassembly with $t$ = 0 being immediately before buffer exchange is initiated. The initial data point for reassembly occurs $\sim$10 min after the final time point of disassembly due to the time needed to switch reservoirs and initiate buffer exchange. We also carried out measurements in steady-state monomer solutions and fully-assembled networks as a control to compare to our initial and final time measurements. Steady-state moduli values are comparable to previously reported values~\cite{ziemann1994local,schmidt2000microrheometry, gardel2003microrheology,schnurr1997determining, wachsstock1994cross} and to our initial and final time measurements (Fig.~S1). 

\textbf{Mathematical Model:} 
To obtain the mechanical response of the network as a function of time, we use an integrated approach where we first use Master Equations to calculate the average filament length of the network
$\langle L \rangle$ as a function of time, and then use Rigidity Percolation theory to obtain the linear storage modulus $G'$ as a function of $\langle L \rangle$. We use these results to obtain $G'$  as a function of time, for both the disassembling and reassembling network. Below, we summarize our mathematical model and method, which is fully described in Supplemental Information. 

Master Equations for Filament Length: We study the time evolution of the concentration of filaments of length $L$, under the influence of microscopic processes that lead to the growth or shrinkage of actin filaments, using a Master Equation framework, described in detail in the SI. For depolymerization, we consider a spatially homogeneous system with a fixed amount of actin, initially present in the form of filaments of a given length in a network.  Assuming that the filaments depolymerize independent of each other, we obtain the probability, $P(L,t)$, that a filament has length $L$ at time $t$, for a given rate, $k_{\rm off}$, of depolymerization. We assume that the polymerization rate is negligibly small during this stage. 

The Master Equation for filament reassembly, on the other hand, is more complex and has two stages: nucleation and polymerization. We assume that at the beginning of reassembly, actin is present in the form of monomers only. The spontaneous nucleation of an actin filament is experimentally known to consist of the formation of a trimeric nucleus. Since the intermediate actin dimer is very unstable, this process is kinetically unfavorable; it is much slower than polymerization of existing actin filaments and is the rate limiting step in actin filament growth \cite{nucleationref1,nucleationref2}. In our simulations, the polymerization rate $k_+$ is varied from 0.1 --  $s^{-1}$ and the depolymerization rate $k_{\rm off}$ is set to 0.16 $s^{-1}$, based on values reported in the literature~\citep{jegou2011individual}. We have considered nucleation rates $k_{\rm nuc} \sim10^{-2}$ -- $10^{-3}$ $s^{-1}$ (see Fig. S5), and the simulation results reported in Fig.~\ref{ffigtheory} were obtained using a nucleation rate of $10^{-3}$ $s^{-1}$. The probability $P(L,t)$ during the filament reassembly phase is obtained by solving the corresponding Master equation as described in the SI, and is used to calculate the average filament length at a given time $t$, $\langle L \rangle = \sum L P(L,t)$. In converting average filament length from monomers to $\mu m$, we assume that each monomer is 6 $nm$ long and steady-state filament lengths are 3 -- 11 $\mu$m~\citep{byrne2014molecules, burlacu1992distribution,gurmessa2016entanglement}. We do not explicitly take into account interactions with the surrounding solvent.  

Rigidity Percolation Theory: The rigidity percolation theory models a biopolymer network as a disordered network made of fibers, and consisting of both flexible (sparsely connected) and rigid (densely connected) regions. When flexible regions span the network, external strains can be accommodated without stretching or bending the fibers. As a result the entire network is flexible, does not resist external strains and has a zero storage modulus. On the other hand, when the rigid regions percolate, the network becomes mechanically rigid, and has a non-zero storage modulus. The system therefore goes from non-rigid to rigid at a certain number density of fibers known as the rigidity percolation threshold, and this mechanical phase transition is known as the rigidity percolation transition. Within the rigid phase, there are two mechanical regimes: an affinely deforming regime corresponding to a dense, rigid network that deforms uniformly at all length scales and has a storage modulus that decreases very slowly with fiber content; and a floppy, barely-rigid network that deforms non-uniformly or non-affinely at different length scales with large variations in the storage modulus with small variations in fiber content \cite{Das1}. Details of this model are described in the SI. The storage modulus of the network $G'$ for a given average fiber length $\langle L \rangle$ is obtained by means of an effective medium theory calculation \cite{Das1, Das2}. 

\section*{Results and Discussion}

We measure the storage and loss moduli, $G'$ and $G''$, of entangled and crosslinked actin networks as a function of time during filament depolymerization and subsequent repolymerization induced by microfluidic-enabled buffer exchange. We focus our discussion on the time evolution of $G'$, as it is a better indicator of network formation than $G''$, since it quantifies the elasticity of a network and becomes essentially negligible for a solution of monomers. Fig.~\ref{ffig1} shows the frequency-averaged storage moduli measured during induced depolymerization~(Fig.~\ref{ffig1}A) and repolymerization~(Fig.~\ref{ffig1}B).

\begin{figure}[ht!]
\centerline{\includegraphics[height=0.6\textwidth]{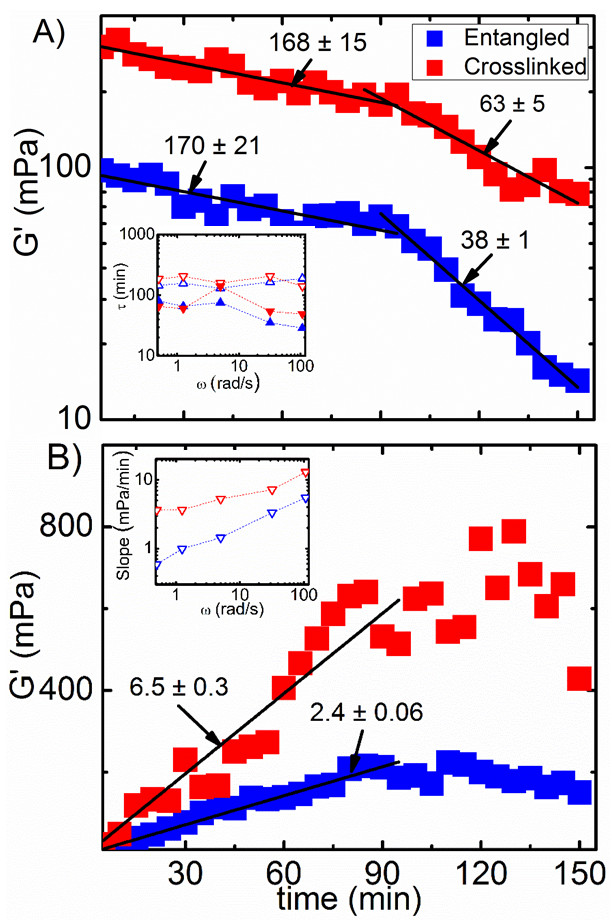}}
\caption{Time course of frequency-averaged storage modulus for entangled (blue) and crosslinked (red)  actin networks during (A) disassembly and (B) reassembly. Solid lines are fits to the data corresponding to, in (A), single exponential decays with time constants of 170 $\pm$ 21 min, 168 $\pm$ 15 min for $t$ $<$ 90 min (phase 1) and 38 $\pm$ 1 min, 63 $\pm$ 5 min for $t$ $>$ 90 min (phase 2); and, in (B), linear functions with slopes 2.4 $\pm$ 0.06 mPa/min and 6.5 $\pm$ 0.3 mPa/min, for entangled and crosslinked networks, respectively. The insets show, in (A), the time constants for phase 1 (open symbols) and phase 2 (closed symbols) and, in (B), the slopes of the fits measured for each oscillation frequency.}\label{ffig1}
\end{figure}

To verify that the systems are fully dis/re-assembling we compare our measured values at the beginning and end of de/re-polymerization to those of steady-state networks and solutions of monomers. As shown in Fig. S1, the moduli measured at the beginning and end of each process are in good agreement with those measured in steady-state. We note that, at the end of triggered disassembly, $G'$ for the crosslinked network indicates that it has not fully disassembled; however, the remaining disassembly completes during the $\sim$8 - 10 min time frame required to begin subsequent reassembly, as indicated by the initial moduli measured for reassembly. 

During depolymerization, we find that the storage moduli of both systems decrease exponentially with time~(Fig.~\ref{ffig1}A). Rather than a single exponential decay, which would indicate a single static process dictating mechanics, we instead find two distinct exponential phases for both entangled and crosslinked systems: initial slow decay of $G'$ for $t$ $<$ 90 min with an average time constant of $\tau_1$ = 169 $\pm$ 26 min, followed by fast decay, with an average time constant of $\tau_2$ = 51 $\pm$ 5 min.  

During reassembly, $G'$ for both systems increases linearly with time for $t$ $<$ 90 min after which they reach steady-state values~(Fig.~\ref{ffig1}B). While the reassembly process appears to be relatively fast ($ \sim$90 min) compared to disassembly ($\sim$150 min), the time needed to reach apparent complete reassembly is nearly identical to the time needed to switch to fast decay during disassembly. This reversible crossover timescale is likely the timescale at which a fully percolated network is achieved or destroyed, which we discuss further below. 

\begin{figure}[ht!]
\centering
\includegraphics[height=0.6\textwidth]{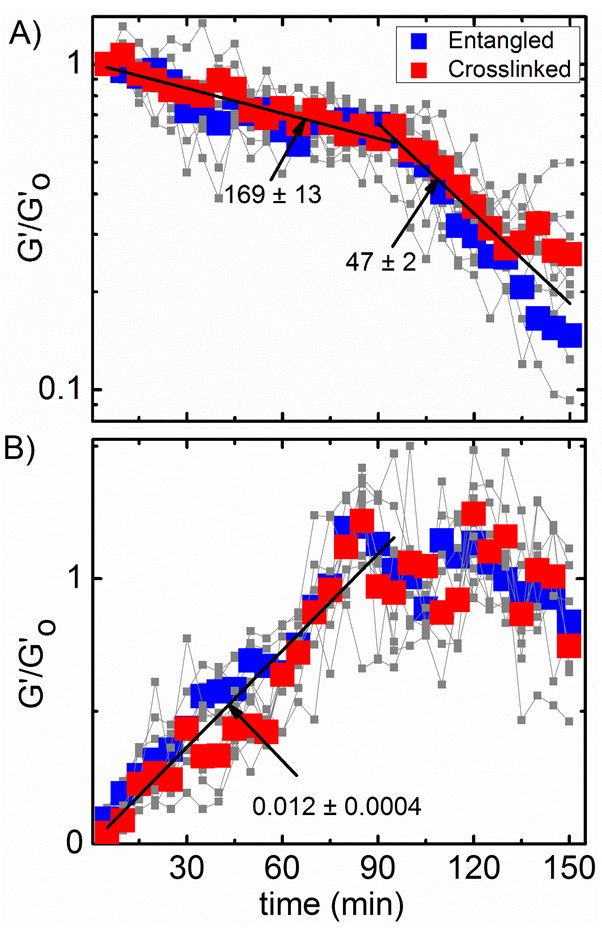}
\caption{Time course of frequency-averaged storage modulus $G'$ normalized by the network modulus $G'_o$ for entangled (blue) and crosslinked (red)  actin networks during (A) disassembly and (B) reassembly. Solid lines are, in (A), fits to the average of the data  for entangled and crosslinked networks, corresponding to single exponential decays with time constants of 169 $\pm$ 13 min for $t$ $<$ 90 min and 47 $\pm$ 2 min for $t$ $>$ 90 min, and, in (B), a linear function with slope 0.012 $\pm$ 0.0004 min $^{-1}$. The grey data are the moduli measured for each frequency.}\label{ffig2}
\end{figure}

As shown in the insets of Fig.~\ref{ffig1}, the dependence of measured decay times (Fig.~\ref{ffig1}A) and linear slopes (Fig.~\ref{ffig1}B) on oscillation frequency are minimal. Further, while the magnitudes of $G'$ are larger for crosslinked networks compared to entangled systems (as expected), the time evolution of $G'$ is remarkably similar for both. To further demonstrate the insensitivity of measured trends on oscillation frequency and network architecture, in Fig.~\ref{ffig2} we plot $G'$ normalized by the corresponding steady-state network value, $G'_o$, for each system and oscillation frequency. As shown, the time evolution of $G'/G'_0$ for all frequencies and both networks collapse to a single master curve suggesting that the molecular mechanisms driving self-assembly of actin networks are universal and do not depend on how the filaments interact or the timescale of mechanical perturbations.

We further note that the time evolution of both $G''$ and the complex viscosity are similar to that for the $G'$ data with comparable decay times (disassembly) and linear slopes (reassembly) (Figs. S2, S3, Tables S1--S3). As with $G'$, these trends are largely independent of oscillation frequency and crosslinking.

\begin{figure}[ht!]
\centerline{\includegraphics[width=0.45\textwidth]{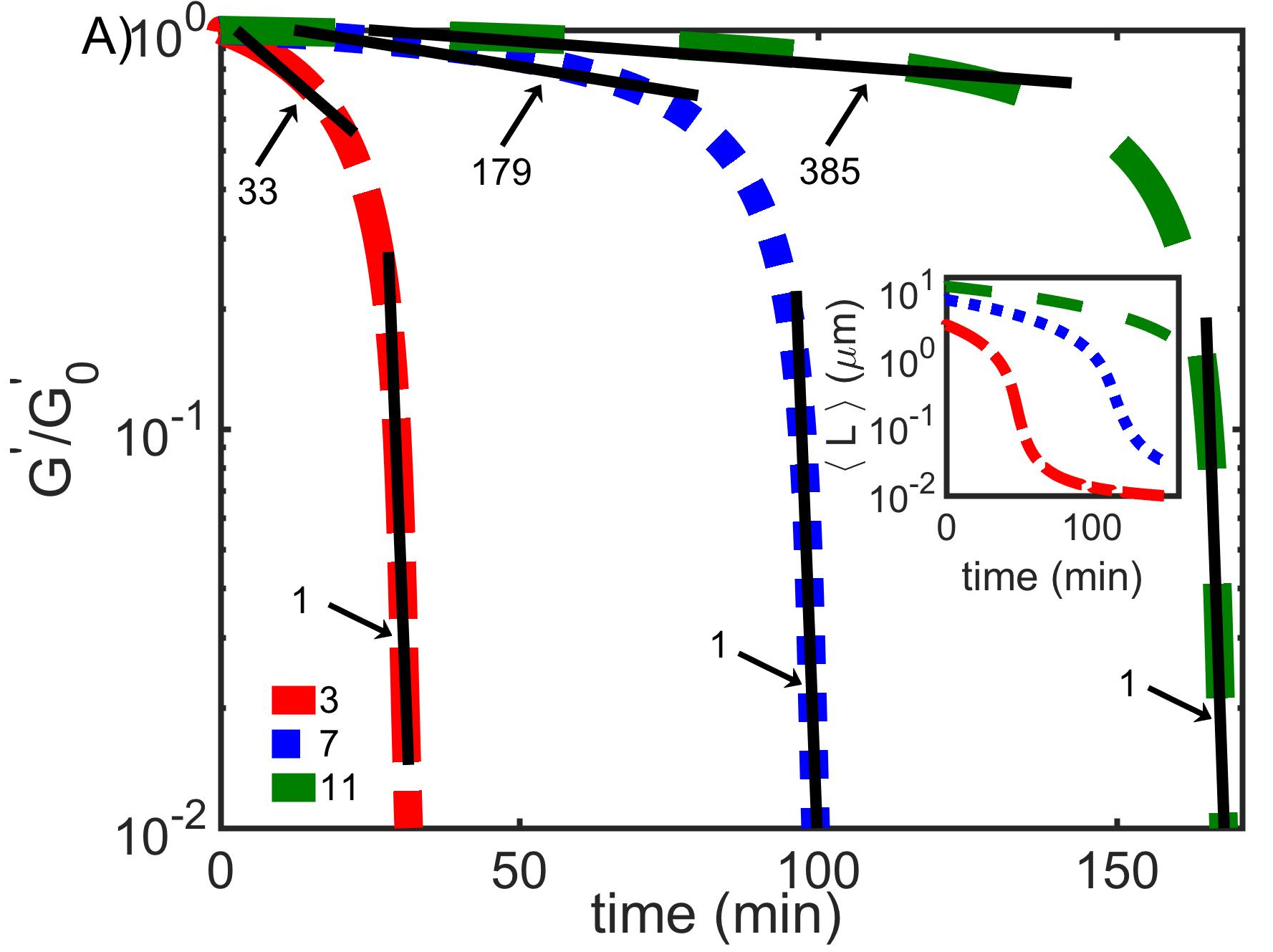}}
\centerline{\includegraphics[width=0.45\textwidth]{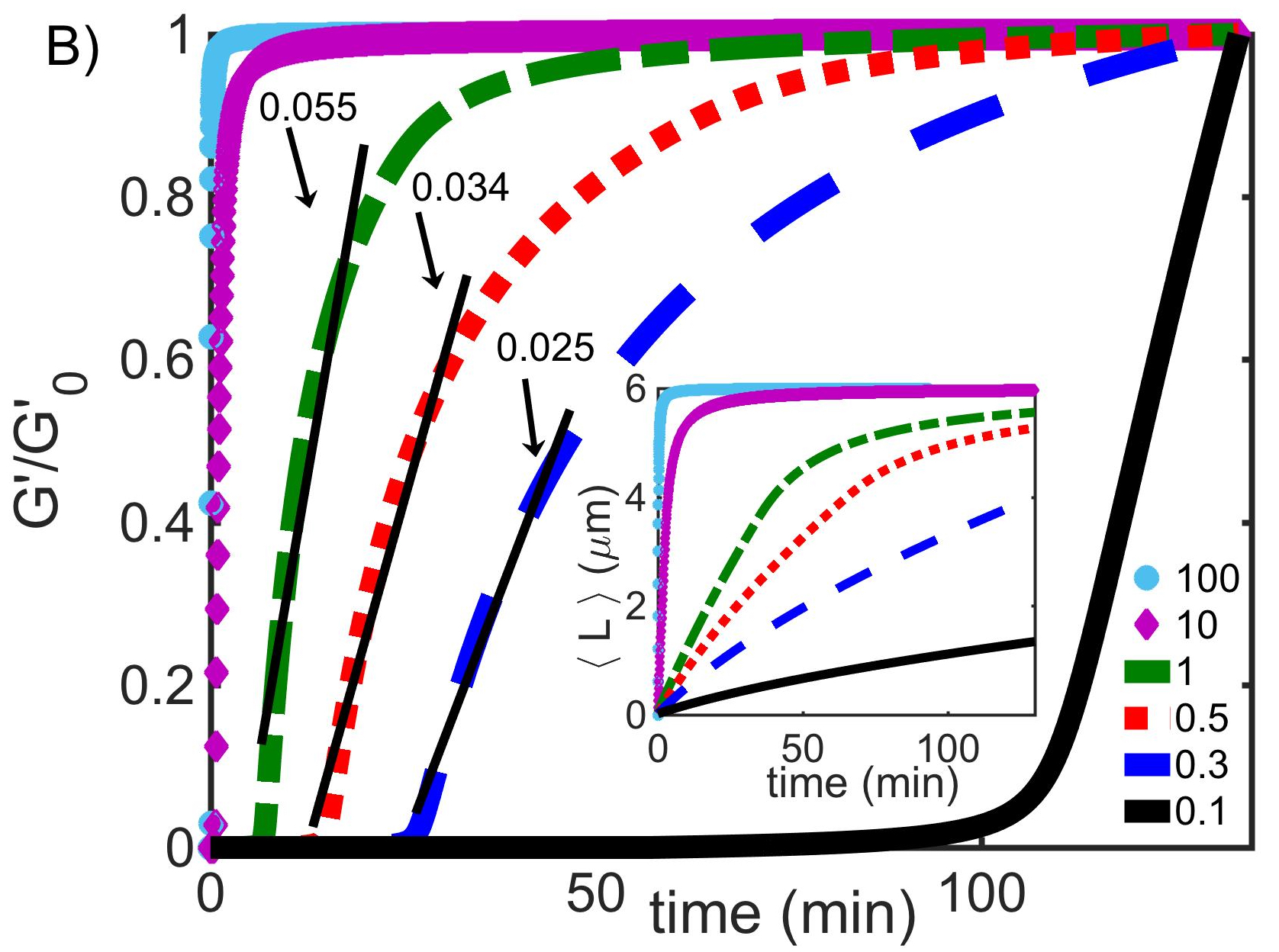}}
\caption{Predicted normalized storage modulus and average filament length (inset) as a function of time during network (A) disassembly and (B) reassembly. For disassembly, the displayed curves correspond to 3 different initial filament lengths, listed in microns in the legend, with the straight solid lines showing exponential fits with decay times displayed in minutes. The depolymerization rate is set to 0.16 $s^{-1}$. For reassembly, displayed curves correspond to  6 different elongation rates, listed in units of $s^{-1}$ in the legend, with solid straight lines showing linear fits with slopes displayed in units of min$^{-1}$. The nucleation rate is set to $0.001$ $s^{-1}$.  Modulus values are normalized by the modulus value for a fully percolated network.} 
\label{ffigtheory}
\end{figure}

In order to relate the mechanical response to molecular processes occurring during dis/re-assembly, we developed a mathematical model, detailed in Methods and SI, that couples the polymerization kinetics of single filaments to the formation and destruction of a rigidly percolated network. For disassembly, we test initial filament lengths that span the experimentally measured distribution of initial lengths~(7.2 $\pm$ 4.0 $\mu$m, Fig. S4). As shown in Fig.~\ref{ffigtheory}A, our model predicts that both the average filament length  $\langle L \rangle$ and the storage modulus $G'$ initially decrease slowly with time before undergoing a distinct crossover to far more rapid decay. Similar to our experimental data, both fast and slow phases of $G'/G'_0$ decay exhibit exponential dependence on time with characteristic time constants. The initial slow time constant as well as the crossover time increase with increasing initial filament length (Fig.~\ref{ffigtheory}A, Table S4). For an initial filament length of 7 ${\mu}$m, corresponding to the experimentally measured average initial filament length, the decay time for the initial slow phase is $\sim$179 min and the crossover time is $\sim$87 min. These theoretical predictions are in excellent agreement with experimentally measured timescales (Figs.~\ref{ffig1}A,~\ref{ffig2}A). Our predicted fast decay time constant ($\tau_2$ $\sim 1 $ min) is, however, an order of magnitude faster than the experimental value. As described below, this difference likely arises from the polydispersity in initial filament lengths in the experiments, whereas the model uses monodisperse initial filament lengths.

The mechanical regime corresponding to the initial slow decay corresponds to an affinely deforming dense network, while the regime corresponding to the sharp decrease corresponds to a non-affinely deforming sparse network. The storage modulus drops to negligibly small values as the network goes through the rigidity percolation phase transition, going from a mechanically rigid regime to a non-rigid regime. Namely, as  the average filament length falls below a certain threshold, the network is no longer able to bear macroscopic stresses and its storage modulus approaches zero. As shown in Fig.~S4, our experimental filament lengths are gamma distributed with an exponential tail of long filaments up to $\sim$20 $\mu$m. These long filaments slow the complete destruction of the network. As such, the second phase in our experimental measurements is a combination of the slow decay from long filaments still in the rigid regime (as demonstrated by the 11 $\mu$m curve in Fig.~\ref{ffigtheory}A) and the fast decay from shorter filaments. Thus, as a result of the inherently heterogeneous and noisy nature of the material in the experiments, the rigidity percolation phase transition is broader and not as sharp as in the model.

During reassembly, our model also predicts a crossover from non-affine to affine mechanical regimes, and varies with the chosen polymerization rate $k_{+}$. The fastest rate we chose ($k_{+}$ = 100 $s^{-1}$) corresponds to the value most often assumed in recent literature~\cite{ pollard1986rate, jegou2011individual,crevenna2015side,fujiwara2007polymerization,fujiwara2002microscopic}, while the slower values (0.1 - 1 $s^{-1}$) are from earlier work~\citep{melki1996continuous, carlier1987mechanisms}. As shown in Fig.~\ref{ffigtheory}B, for polymerization rates of 100, 10, 1, 0.5, 0.3, and 0.1 $s^{-1}$, the crossover takes place at $\sim$2, 5, 25, 40, 60, and 130 min, respectively. The corresponding initial slopes of $G'/G'_0$ curves increase from $\sim$0.025 to 1.791 min$^{-1}$ as $k_{+}$ increases from 0.3 to 100 $s^{-1}$ (Table S5). The predicted crossover time and slope closest to our experimental observations (within a factor $\sim$2) are for the relatively slow rate of $k_{+}$ $\sim$ 0.3 $s^{-1}$, whereas the crossover time and slope for the commonly assumed rate of $k_{+} \sim$ 100 $s^{-1}$ differ by nearly two orders of magnitude from experimental values. We note that the crossover time and slope also change with nucleation rate, but this is a much smaller effect in comparison (Fig. S5). These results indicate that elongation kinetics and the mechanical phase transition in network rigidity dominate the mechanical response of evolving networks. 

\section*{Conclusion}
 We have presented an experimental approach that combines time-resolved optical tweezers microrheology with diffusion-controlled microfluidics, to measure the time-evolution of microscale mechanical properties of dynamic systems during triggered activity. We use this technique to measure the viscoelastic moduli of actin networks during disassembly and reassembly via chemically-induced de/re-polymerization of actin filaments. To inform our experimental results, we develop an integrated mathematical model that couples the time-evolution of filament lengths with rigidity percolation theory to predict the evolution of the storage modulus during network disassembly and reassembly. We find that actin networks exhibit biphasic exponential decay with time during network disassembly, which we show arises from a rigidity percolation phase transition at $\sim$90 min as the network is destroyed (Figs.~\ref{ffig1}A--\ref{ffigtheory}A). During reassembly, the moduli increase linearly for the first $\sim$90 min after which steady-state mechanics is achieved~(Figs.~\ref{ffig1}B,~\ref{ffig2}B). We show that the changing mechanical properties during reassembly can be primarily attributed to the elongation kinetics and rigidity percolation transition of the networks~(Figs.~\ref{ffigtheory}B, S5). Further, elongation rates appear to be much slower than those assumed in recent literature, but in agreement with those from earlier work. Finally, the time-evolution of moduli were markedly similar for both entangled and crosslinked networks and for all oscillation frequencies, demonstrating the universality of the measured mechanics~(Fig.~\ref{ffig2}, Tables S1--S3). Our collective results fill a long-standing gap in knowledge regarding how de/re-polymerization kinetics of cytoskeletal proteins map to time-varying mechanical properties of the network. These results are essential to understanding how cells can dynamically alter their mechanical properties to enable key processes such as motility, division, growth, and apoptosis; and shed light onto the underlying principles of other dynamic, self-assembling systems and materials. Our technique can also be used to measure the time-dependent mechanical properties of the growing number of active materials currently under intense investigation. 
 
 SI file can be accessed at:\\
 https://drive.google.com/file/d/0B65f6vTXpwK4UFpIR1FIcXVCNkE/view?usp=sharing 
 
\begin{acknowledgement}
This research was funded by an NSF CAREER Award, grant number 1255446 to RMR-A, and Research Corporation \& Gordon \& Betty Moore Foundation Collaborative Innovation Award to RMR-A and JLR. MD and NB were partially supported by a Cottrell College Science Award from Research Corporation. MD further acknowledges the hospitality of the Aspen Center for Physics. 
\end{acknowledgement}

\bibliography{Manuscript.bbl}

\providecommand{\latin}[1]{#1}
\makeatletter
\providecommand{\doi}
  {\begingroup\let\do\@makeother\dospecials
  \catcode`\{=1 \catcode`\}=2 \doi@aux}
\providecommand{\doi@aux}[1]{\endgroup\texttt{#1}}
\makeatother
\providecommand*\mcitethebibliography{\thebibliography}
\csname @ifundefined\endcsname{endmcitethebibliography}
  {\let\endmcitethebibliography\endthebibliography}{}
\begin{mcitethebibliography}{54}
\providecommand*\natexlab[1]{#1}
\providecommand*\mciteSetBstSublistMode[1]{}
\providecommand*\mciteSetBstMaxWidthForm[2]{}
\providecommand*\mciteBstWouldAddEndPuncttrue
  {\def\EndOfBibitem{\unskip.}}
\providecommand*\mciteBstWouldAddEndPunctfalse
  {\let\EndOfBibitem\relax}
\providecommand*\mciteSetBstMidEndSepPunct[3]{}
\providecommand*\mciteSetBstSublistLabelBeginEnd[3]{}
\providecommand*\EndOfBibitem{}
\mciteSetBstSublistMode{f}
\mciteSetBstMaxWidthForm{subitem}{(\alph{mcitesubitemcount})}
\mciteSetBstSublistLabelBeginEnd
  {\mcitemaxwidthsubitemform\space}
  {\relax}
  {\relax}

\bibitem[Paavilainen \latin{et~al.}(2004)Paavilainen, Bertling, Falck, and
  Lappalainen]{paavilainen2004regulation}
Paavilainen,~V.~O.; Bertling,~E.; Falck,~S.; Lappalainen,~P. Regulation of
  cytoskeletal dynamics by actin-monomer-binding proteins. \emph{Trends in cell
  biology} \textbf{2004}, \emph{14}, 386--394\relax
\mciteBstWouldAddEndPuncttrue
\mciteSetBstMidEndSepPunct{\mcitedefaultmidpunct}
{\mcitedefaultendpunct}{\mcitedefaultseppunct}\relax
\EndOfBibitem
\bibitem[Stricker \latin{et~al.}(2010)Stricker, Falzone, and
  Gardel]{stricker2010mechanics}
Stricker,~J.; Falzone,~T.; Gardel,~M.~L. Mechanics of the F-actin cytoskeleton.
  \emph{Journal of biomechanics} \textbf{2010}, \emph{43}, 9--14\relax
\mciteBstWouldAddEndPuncttrue
\mciteSetBstMidEndSepPunct{\mcitedefaultmidpunct}
{\mcitedefaultendpunct}{\mcitedefaultseppunct}\relax
\EndOfBibitem
\bibitem[Wen and Janmey(2011)Wen, and Janmey]{wen2011polymer}
Wen,~Q.; Janmey,~P.~A. Polymer physics of the cytoskeleton. \emph{Current
  Opinion in Solid State and Materials Science} \textbf{2011}, \emph{15},
  177--182\relax
\mciteBstWouldAddEndPuncttrue
\mciteSetBstMidEndSepPunct{\mcitedefaultmidpunct}
{\mcitedefaultendpunct}{\mcitedefaultseppunct}\relax
\EndOfBibitem
\bibitem[Gardel \latin{et~al.}(2008)Gardel, Kasza, Brangwynne, Liu, and
  Weitz]{gardel2008mechanical}
Gardel,~M.~L.; Kasza,~K.~E.; Brangwynne,~C.~P.; Liu,~J.; Weitz,~D.~A.
  Mechanical response of cytoskeletal networks. \emph{Methods in cell biology}
  \textbf{2008}, \emph{89}, 487--519\relax
\mciteBstWouldAddEndPuncttrue
\mciteSetBstMidEndSepPunct{\mcitedefaultmidpunct}
{\mcitedefaultendpunct}{\mcitedefaultseppunct}\relax
\EndOfBibitem
\bibitem[Cooper and Ganem(1997)Cooper, and Ganem]{cooper1997cell}
Cooper,~G.~M.; Ganem,~D. The Cell: A Molecular Approach. \emph{Nature Medicine}
  \textbf{1997}, \emph{3}, 1042--1042\relax
\mciteBstWouldAddEndPuncttrue
\mciteSetBstMidEndSepPunct{\mcitedefaultmidpunct}
{\mcitedefaultendpunct}{\mcitedefaultseppunct}\relax
\EndOfBibitem
\bibitem[Byrne \latin{et~al.}(2014)Byrne, Heidelberger, and
  Waxham]{byrne2014molecules}
Byrne,~J.~H.; Heidelberger,~R.; Waxham,~M.~N. \emph{From molecules to networks:
  an introduction to cellular and molecular neuroscience}; Academic Press,
  2014\relax
\mciteBstWouldAddEndPuncttrue
\mciteSetBstMidEndSepPunct{\mcitedefaultmidpunct}
{\mcitedefaultendpunct}{\mcitedefaultseppunct}\relax
\EndOfBibitem
\bibitem[Lieleg \latin{et~al.}(2008)Lieleg, Claessens, Luan, and
  Bausch]{lieleg2008transient}
Lieleg,~O.; Claessens,~M.; Luan,~Y.; Bausch,~A. Transient binding and
  dissipation in cross-linked actin networks. \emph{Physical review letters}
  \textbf{2008}, \emph{101}, 108101\relax
\mciteBstWouldAddEndPuncttrue
\mciteSetBstMidEndSepPunct{\mcitedefaultmidpunct}
{\mcitedefaultendpunct}{\mcitedefaultseppunct}\relax
\EndOfBibitem
\bibitem[Luan \latin{et~al.}(2008)Luan, Lieleg, Wagner, and
  Bausch]{luan2008micro}
Luan,~Y.; Lieleg,~O.; Wagner,~B.; Bausch,~A.~R. Micro-and macrorheological
  properties of isotropically cross-linked actin networks. \emph{Biophysical
  journal} \textbf{2008}, \emph{94}, 688--693\relax
\mciteBstWouldAddEndPuncttrue
\mciteSetBstMidEndSepPunct{\mcitedefaultmidpunct}
{\mcitedefaultendpunct}{\mcitedefaultseppunct}\relax
\EndOfBibitem
\bibitem[Tellam(1985)]{tellam1985mechanism}
Tellam,~R. Mechanism of calcium chloride induced actin polymerization.
  \emph{Biochemistry} \textbf{1985}, \emph{24}, 4455--4460\relax
\mciteBstWouldAddEndPuncttrue
\mciteSetBstMidEndSepPunct{\mcitedefaultmidpunct}
{\mcitedefaultendpunct}{\mcitedefaultseppunct}\relax
\EndOfBibitem
\bibitem[Blanchoin and Pollard(2002)Blanchoin, and
  Pollard]{blanchoin2002hydrolysis}
Blanchoin,~L.; Pollard,~T.~D. Hydrolysis of ATP by polymerized actin depends on
  the bound divalent cation but not profilin. \emph{Biochemistry}
  \textbf{2002}, \emph{41}, 597--602\relax
\mciteBstWouldAddEndPuncttrue
\mciteSetBstMidEndSepPunct{\mcitedefaultmidpunct}
{\mcitedefaultendpunct}{\mcitedefaultseppunct}\relax
\EndOfBibitem
\bibitem[Fujiwara \latin{et~al.}(2002)Fujiwara, Takahashi, Tadakuma, Funatsu,
  and Ishiwata]{fujiwara2002microscopic}
Fujiwara,~I.; Takahashi,~S.; Tadakuma,~H.; Funatsu,~T.; Ishiwata,~S.
  Microscopic analysis of polymerization dynamics with individual actin
  filaments. \emph{Nature cell biology} \textbf{2002}, \emph{4}, 666--673\relax
\mciteBstWouldAddEndPuncttrue
\mciteSetBstMidEndSepPunct{\mcitedefaultmidpunct}
{\mcitedefaultendpunct}{\mcitedefaultseppunct}\relax
\EndOfBibitem
\bibitem[Pollard(1986)]{pollard1986rate}
Pollard,~T.~D. Rate constants for the reactions of ATP-and ADP-actin with the
  ends of actin filaments. \emph{The Journal of cell biology} \textbf{1986},
  \emph{103}, 2747--2754\relax
\mciteBstWouldAddEndPuncttrue
\mciteSetBstMidEndSepPunct{\mcitedefaultmidpunct}
{\mcitedefaultendpunct}{\mcitedefaultseppunct}\relax
\EndOfBibitem
\bibitem[Li \latin{et~al.}(2009)Li, Kierfeld, and Lipowsky]{li2009actin}
Li,~X.; Kierfeld,~J.; Lipowsky,~R. Actin polymerization and depolymerization
  coupled to cooperative hydrolysis. \emph{Physical review letters}
  \textbf{2009}, \emph{103}, 048102\relax
\mciteBstWouldAddEndPuncttrue
\mciteSetBstMidEndSepPunct{\mcitedefaultmidpunct}
{\mcitedefaultendpunct}{\mcitedefaultseppunct}\relax
\EndOfBibitem
\bibitem[Amann and Pollard(2001)Amann, and Pollard]{amann2001direct}
Amann,~K.~J.; Pollard,~T.~D. Direct real-time observation of actin filament
  branching mediated by Arp2/3 complex using total internal reflection
  fluorescence microscopy. \emph{Proceedings of the National Academy of
  Sciences} \textbf{2001}, \emph{98}, 15009--15013\relax
\mciteBstWouldAddEndPuncttrue
\mciteSetBstMidEndSepPunct{\mcitedefaultmidpunct}
{\mcitedefaultendpunct}{\mcitedefaultseppunct}\relax
\EndOfBibitem
\bibitem[Melki \latin{et~al.}(1996)Melki, Fievez, and
  Carlier]{melki1996continuous}
Melki,~R.; Fievez,~S.; Carlier,~M.-F. Continuous monitoring of Pi release
  following nucleotide hydrolysis in actin or tubulin assembly using
  2-amino-6-mercapto-7-methylpurine ribonucleoside and purine-nucleoside
  phosphorylase as an enzyme-linked assay. \emph{Biochemistry} \textbf{1996},
  \emph{35}, 12038--12045\relax
\mciteBstWouldAddEndPuncttrue
\mciteSetBstMidEndSepPunct{\mcitedefaultmidpunct}
{\mcitedefaultendpunct}{\mcitedefaultseppunct}\relax
\EndOfBibitem
\bibitem[Carlier and Pantaloni(1986)Carlier, and Pantaloni]{carlier1986direct}
Carlier,~M.; Pantaloni,~D. Direct evidence for ADP-inorganic phosphate-F-actin
  as the major intermediate in ATP-actin polymerization. Rate of dissociation
  of inorganic phosphate from actin filaments. \emph{Biochemistry}
  \textbf{1986}, \emph{25}, 7789--7792\relax
\mciteBstWouldAddEndPuncttrue
\mciteSetBstMidEndSepPunct{\mcitedefaultmidpunct}
{\mcitedefaultendpunct}{\mcitedefaultseppunct}\relax
\EndOfBibitem
\bibitem[Fujiwara \latin{et~al.}(2007)Fujiwara, Vavylonis, and
  Pollard]{fujiwara2007polymerization}
Fujiwara,~I.; Vavylonis,~D.; Pollard,~T.~D. Polymerization kinetics of ADP-and
  ADP-Pi-actin determined by fluorescence microscopy. \emph{Proceedings of the
  National Academy of Sciences} \textbf{2007}, \emph{104}, 8827--8832\relax
\mciteBstWouldAddEndPuncttrue
\mciteSetBstMidEndSepPunct{\mcitedefaultmidpunct}
{\mcitedefaultendpunct}{\mcitedefaultseppunct}\relax
\EndOfBibitem
\bibitem[J{\'e}gou \latin{et~al.}(2011)J{\'e}gou, Niedermayer, Orb{\'a}n,
  Didry, Lipowsky, Carlier, and Romet-Lemonne]{jegou2011individual}
J{\'e}gou,~A.; Niedermayer,~T.; Orb{\'a}n,~J.; Didry,~D.; Lipowsky,~R.;
  Carlier,~M.-F.; Romet-Lemonne,~G. Individual actin filaments in a
  microfluidic flow reveal the mechanism of ATP hydrolysis and give insight
  into the properties of profilin. \emph{PLoS Biol} \textbf{2011}, \emph{9},
  e1001161\relax
\mciteBstWouldAddEndPuncttrue
\mciteSetBstMidEndSepPunct{\mcitedefaultmidpunct}
{\mcitedefaultendpunct}{\mcitedefaultseppunct}\relax
\EndOfBibitem
\bibitem[Crevenna \latin{et~al.}(2015)Crevenna, Arciniega, Dupont, Mizuno,
  Kowalska, Lange, Wedlich-S{\"o}ldner, and Lamb]{crevenna2015side}
Crevenna,~A.~H.; Arciniega,~M.; Dupont,~A.; Mizuno,~N.; Kowalska,~K.;
  Lange,~O.~F.; Wedlich-S{\"o}ldner,~R.; Lamb,~D.~C. Side-binding proteins
  modulate actin filament dynamics. \emph{Elife} \textbf{2015}, \emph{4},
  e04599\relax
\mciteBstWouldAddEndPuncttrue
\mciteSetBstMidEndSepPunct{\mcitedefaultmidpunct}
{\mcitedefaultendpunct}{\mcitedefaultseppunct}\relax
\EndOfBibitem
\bibitem[Janmey \latin{et~al.}(1994)Janmey, Hvidt, K{\"a}s, Lerche, Maggs,
  Sackmann, Schliwa, and Stossel]{janmey1994mechanical}
Janmey,~P.~A.; Hvidt,~S.; K{\"a}s,~J.; Lerche,~D.; Maggs,~A.; Sackmann,~E.;
  Schliwa,~M.; Stossel,~T.~P. The mechanical properties of actin gels. Elastic
  modulus and filament motions. \emph{Journal of Biological Chemistry}
  \textbf{1994}, \emph{269}, 32503--32513\relax
\mciteBstWouldAddEndPuncttrue
\mciteSetBstMidEndSepPunct{\mcitedefaultmidpunct}
{\mcitedefaultendpunct}{\mcitedefaultseppunct}\relax
\EndOfBibitem
\bibitem[Janmey \latin{et~al.}(1988)Janmey, Hvidt, Peetermans, Lamb, Ferry, and
  Stossel]{janmey1988viscoelasticity}
Janmey,~P.~A.; Hvidt,~S.; Peetermans,~J.; Lamb,~J.; Ferry,~J.~D.;
  Stossel,~T.~P. Viscoelasticity of F-actin and F-actin/gelsolin complexes.
  \emph{Biochemistry} \textbf{1988}, \emph{27}, 8218--8227\relax
\mciteBstWouldAddEndPuncttrue
\mciteSetBstMidEndSepPunct{\mcitedefaultmidpunct}
{\mcitedefaultendpunct}{\mcitedefaultseppunct}\relax
\EndOfBibitem
\bibitem[Kueh \latin{et~al.}(2008)Kueh, Brieher, and
  Mitchison]{kueh2008dynamic}
Kueh,~H.~Y.; Brieher,~W.~M.; Mitchison,~T.~J. Dynamic stabilization of actin
  filaments. \emph{Proceedings of the National Academy of Sciences}
  \textbf{2008}, \emph{105}, 16531--16536\relax
\mciteBstWouldAddEndPuncttrue
\mciteSetBstMidEndSepPunct{\mcitedefaultmidpunct}
{\mcitedefaultendpunct}{\mcitedefaultseppunct}\relax
\EndOfBibitem
\bibitem[Zaner(1995)]{zaner1995physics}
Zaner,~K.~S. Physics of actin networks. I. Rheology of semi-dilute F-actin.
  \emph{Biophysical journal} \textbf{1995}, \emph{68}, 1019--1026\relax
\mciteBstWouldAddEndPuncttrue
\mciteSetBstMidEndSepPunct{\mcitedefaultmidpunct}
{\mcitedefaultendpunct}{\mcitedefaultseppunct}\relax
\EndOfBibitem
\bibitem[Xu \latin{et~al.}(1998)Xu, Palmer, and Wirtz]{xu1998rheology}
Xu,~J.; Palmer,~A.; Wirtz,~D. Rheology and microrheology of semiflexible
  polymer solutions: actin filament networks. \emph{Macromolecules}
  \textbf{1998}, \emph{31}, 6486--6492\relax
\mciteBstWouldAddEndPuncttrue
\mciteSetBstMidEndSepPunct{\mcitedefaultmidpunct}
{\mcitedefaultendpunct}{\mcitedefaultseppunct}\relax
\EndOfBibitem
\bibitem[Ziemann \latin{et~al.}(1994)Ziemann, R{\"a}dler, and
  Sackmann]{ziemann1994local}
Ziemann,~F.; R{\"a}dler,~J.; Sackmann,~E. Local measurements of viscoelastic
  moduli of entangled actin networks using an oscillating magnetic bead
  micro-rheometer. \emph{Biophysical Journal} \textbf{1994}, \emph{66},
  2210--2216\relax
\mciteBstWouldAddEndPuncttrue
\mciteSetBstMidEndSepPunct{\mcitedefaultmidpunct}
{\mcitedefaultendpunct}{\mcitedefaultseppunct}\relax
\EndOfBibitem
\bibitem[Frey \latin{et~al.}(1999)Frey, Kroy, and
  Wilhelm]{frey1999viscoelasticity}
Frey,~E.; Kroy,~K.; Wilhelm,~J. Viscoelasticity of biopolymer networks and
  statistical mechanics of semiflexible polymers. \emph{Advances in structural
  biology} \textbf{1999}, \emph{5}, 135--168\relax
\mciteBstWouldAddEndPuncttrue
\mciteSetBstMidEndSepPunct{\mcitedefaultmidpunct}
{\mcitedefaultendpunct}{\mcitedefaultseppunct}\relax
\EndOfBibitem
\bibitem[Brau \latin{et~al.}(2007)Brau, Ferrer, Lee, Castro, Tam, Tarsa,
  Matsudaira, Boyce, Kamm, and Lang]{brau2007passive}
Brau,~R.; Ferrer,~J.; Lee,~H.; Castro,~C.; Tam,~B.; Tarsa,~P.; Matsudaira,~P.;
  Boyce,~M.; Kamm,~R.; Lang,~M. Passive and active microrheology with optical
  tweezers. \emph{Journal of Optics A: Pure and Applied Optics} \textbf{2007},
  \emph{9}, S103\relax
\mciteBstWouldAddEndPuncttrue
\mciteSetBstMidEndSepPunct{\mcitedefaultmidpunct}
{\mcitedefaultendpunct}{\mcitedefaultseppunct}\relax
\EndOfBibitem
\bibitem[Lee \latin{et~al.}(2010)Lee, Ferrer, Nakamura, Lang, and
  Kamm]{lee2010passive}
Lee,~H.; Ferrer,~J.~M.; Nakamura,~F.; Lang,~M.~J.; Kamm,~R.~D. Passive and
  active microrheology for cross-linked F-actin networks in vitro. \emph{Acta
  biomaterialia} \textbf{2010}, \emph{6}, 1207--1218\relax
\mciteBstWouldAddEndPuncttrue
\mciteSetBstMidEndSepPunct{\mcitedefaultmidpunct}
{\mcitedefaultendpunct}{\mcitedefaultseppunct}\relax
\EndOfBibitem
\bibitem[Morse(1998)]{morse1998viscoelasticity}
Morse,~D.~C. Viscoelasticity of tightly entangled solutions of semiflexible
  polymers. \emph{Physical Review E} \textbf{1998}, \emph{58}, R1237\relax
\mciteBstWouldAddEndPuncttrue
\mciteSetBstMidEndSepPunct{\mcitedefaultmidpunct}
{\mcitedefaultendpunct}{\mcitedefaultseppunct}\relax
\EndOfBibitem
\bibitem[Schmidt \latin{et~al.}(2000)Schmidt, Hinner, and
  Sackmann]{schmidt2000microrheometry}
Schmidt,~F.~G.; Hinner,~B.; Sackmann,~E. Microrheometry underestimates the
  values of the viscoelastic moduli in measurements on F-actin solutions
  compared to macrorheometry. \emph{Physical Review E} \textbf{2000},
  \emph{61}, 5646\relax
\mciteBstWouldAddEndPuncttrue
\mciteSetBstMidEndSepPunct{\mcitedefaultmidpunct}
{\mcitedefaultendpunct}{\mcitedefaultseppunct}\relax
\EndOfBibitem
\bibitem[Kavanagh and Ross-Murphy(1998)Kavanagh, and
  Ross-Murphy]{kavanagh1998rheological}
Kavanagh,~G.~M.; Ross-Murphy,~S.~B. Rheological characterisation of polymer
  gels. \emph{Progress in Polymer Science} \textbf{1998}, \emph{23},
  533--562\relax
\mciteBstWouldAddEndPuncttrue
\mciteSetBstMidEndSepPunct{\mcitedefaultmidpunct}
{\mcitedefaultendpunct}{\mcitedefaultseppunct}\relax
\EndOfBibitem
\bibitem[Gardel \latin{et~al.}(2003)Gardel, Valentine, Crocker, Bausch, and
  Weitz]{gardel2003microrheology}
Gardel,~M.; Valentine,~M.; Crocker,~J.~C.; Bausch,~A.; Weitz,~D. Microrheology
  of entangled F-actin solutions. \emph{Physical review letters} \textbf{2003},
  \emph{91}, 158302\relax
\mciteBstWouldAddEndPuncttrue
\mciteSetBstMidEndSepPunct{\mcitedefaultmidpunct}
{\mcitedefaultendpunct}{\mcitedefaultseppunct}\relax
\EndOfBibitem
\bibitem[Deshpande and Pfohl(2015)Deshpande, and Pfohl]{deshpande2015real}
Deshpande,~S.; Pfohl,~T. Real-time dynamics of emerging actin networks in
  cell-mimicking compartments. \emph{PloS one} \textbf{2015}, \emph{10},
  e0116521\relax
\mciteBstWouldAddEndPuncttrue
\mciteSetBstMidEndSepPunct{\mcitedefaultmidpunct}
{\mcitedefaultendpunct}{\mcitedefaultseppunct}\relax
\EndOfBibitem
\bibitem[Oosawa and Asakura(1975)Oosawa, and Asakura]{Oosawa}
Oosawa,~F.; Asakura,~S. \emph{Thermodynamics of the polymerization of protein /
  Fumio Oosawa and Sho Asakura}; Academic Press London ; New York, 1975; pp
  viii, 204 p. :\relax
\mciteBstWouldAddEndPuncttrue
\mciteSetBstMidEndSepPunct{\mcitedefaultmidpunct}
{\mcitedefaultendpunct}{\mcitedefaultseppunct}\relax
\EndOfBibitem
\bibitem[Sept \latin{et~al.}(1999)Sept, Xu, Pollard, and
  McCammon]{sept1999annealing}
Sept,~D.; Xu,~J.; Pollard,~T.~D.; McCammon,~J.~A. Annealing accounts for the
  length of actin filaments formed by spontaneous polymerization.
  \emph{Biophysical journal} \textbf{1999}, \emph{77}, 2911--2919\relax
\mciteBstWouldAddEndPuncttrue
\mciteSetBstMidEndSepPunct{\mcitedefaultmidpunct}
{\mcitedefaultendpunct}{\mcitedefaultseppunct}\relax
\EndOfBibitem
\bibitem[Mohapatra \latin{et~al.}(2016)Mohapatra, Goode, Jelenkovic, Phillips,
  and Kondev]{ME1}
Mohapatra,~L.; Goode,~B.~L.; Jelenkovic,~P.; Phillips,~R.; Kondev,~J. Design
  Principles of Length Control of Cytoskeletal Structures. \emph{Annual Review
  of Biophysics} \textbf{2016}, \emph{45}, 85--116, DOI:
  \doi{10.1146/annurev-biophys-070915-094206}\relax
\mciteBstWouldAddEndPuncttrue
\mciteSetBstMidEndSepPunct{\mcitedefaultmidpunct}
{\mcitedefaultendpunct}{\mcitedefaultseppunct}\relax
\EndOfBibitem
\bibitem[Tobacman and Korn(1983)Tobacman, and Korn]{ME2}
Tobacman,~L.~S.; Korn,~E.~D. The kinetics of actin nucleation and
  polymerization. \emph{J. Biol. Chem.} \textbf{1983}, \emph{258},
  3207--3214\relax
\mciteBstWouldAddEndPuncttrue
\mciteSetBstMidEndSepPunct{\mcitedefaultmidpunct}
{\mcitedefaultendpunct}{\mcitedefaultseppunct}\relax
\EndOfBibitem
\bibitem[Michaels and Knowles(2014)Michaels, and Knowles]{ME3}
Michaels,~T. C.~T.; Knowles,~T. P.~J. Mean-field master equation formalism for
  biofilament growth. \emph{American Journal of Physics} \textbf{2014},
  \emph{82}, 476--483, DOI: \doi{10.1119/1.4870004}\relax
\mciteBstWouldAddEndPuncttrue
\mciteSetBstMidEndSepPunct{\mcitedefaultmidpunct}
{\mcitedefaultendpunct}{\mcitedefaultseppunct}\relax
\EndOfBibitem
\bibitem[Das \latin{et~al.}(2007)Das, Mackintosh, and Levine]{Das1}
Das,~M.; Mackintosh,~F.~C.; Levine,~A.~J. Effective medium theory of
  semiflexible filamentous networks. \emph{Phys. Rev. Lett.} \textbf{2007},
  \emph{99}, 038101--1\relax
\mciteBstWouldAddEndPuncttrue
\mciteSetBstMidEndSepPunct{\mcitedefaultmidpunct}
{\mcitedefaultendpunct}{\mcitedefaultseppunct}\relax
\EndOfBibitem
\bibitem[Das \latin{et~al.}(2012)Das, Quint, and Schwarz]{Das2}
Das,~M.; Quint,~D.~A.; Schwarz,~J.~M. {R}edundancy and cooperativity in the
  mechanics of compositely crosslinked filamentous networks. \emph{{PL}o{S}
  {ONE}} \textbf{2012}, \emph{7}, e35939--1--11\relax
\mciteBstWouldAddEndPuncttrue
\mciteSetBstMidEndSepPunct{\mcitedefaultmidpunct}
{\mcitedefaultendpunct}{\mcitedefaultseppunct}\relax
\EndOfBibitem
\bibitem[Silverberg \latin{et~al.}(2014)Silverberg, Barrett, Das, Petersen,
  Bonassar, and Cohen]{Das3}
Silverberg,~J.~L.; Barrett,~A.~R.; Das,~M.; Petersen,~P.~B.; Bonassar,~L.~J.;
  Cohen,~I. Structure-Function Relations and Rigidity Percolation in the Shear
  Properties of Articular Cartilage. \emph{Biophysical journal} \textbf{2014},
  \emph{107}, 1721--1730\relax
\mciteBstWouldAddEndPuncttrue
\mciteSetBstMidEndSepPunct{\mcitedefaultmidpunct}
{\mcitedefaultendpunct}{\mcitedefaultseppunct}\relax
\EndOfBibitem
\bibitem[Broedersz \latin{et~al.}(2011)Broedersz, Mao, Lubensky, and
  MacKintosh]{Broedersz}
Broedersz,~C.; Mao,~X.; Lubensky,~T.; MacKintosh,~F.~C. Criticality and
  isostaticity in fibre networks. \emph{Nature Phys.} \textbf{2011}, \emph{7},
  983--988, DOI: \doi{10.1038/nphys2127}\relax
\mciteBstWouldAddEndPuncttrue
\mciteSetBstMidEndSepPunct{\mcitedefaultmidpunct}
{\mcitedefaultendpunct}{\mcitedefaultseppunct}\relax
\EndOfBibitem
\bibitem[Gurmessa \latin{et~al.}(2017)Gurmessa, Ricketts, and
  Robertson-Anderson]{gurmessa2017nonlinear}
Gurmessa,~B.; Ricketts,~S.; Robertson-Anderson,~R.~M. Nonlinear Actin
  Deformations Lead to Network Stiffening, Yielding, and Nonuniform Stress
  Propagation. \emph{Biophysical Journal} \textbf{2017}, \relax
\mciteBstWouldAddEndPunctfalse
\mciteSetBstMidEndSepPunct{\mcitedefaultmidpunct}
{}{\mcitedefaultseppunct}\relax
\EndOfBibitem
\bibitem[Isambert and Maggs(1996)Isambert, and Maggs]{isambert1996dynamics}
Isambert,~H.; Maggs,~A. Dynamics and rheology of actin solutions.
  \emph{Macromolecules} \textbf{1996}, \emph{29}, 1036--1040\relax
\mciteBstWouldAddEndPuncttrue
\mciteSetBstMidEndSepPunct{\mcitedefaultmidpunct}
{\mcitedefaultendpunct}{\mcitedefaultseppunct}\relax
\EndOfBibitem
\bibitem[Park \latin{et~al.}(2016)Park, Jacobson, Nguyen, Willardson, and
  Saleh]{park2016thin}
Park,~C.-Y.; Jacobson,~D.~R.; Nguyen,~D.~T.; Willardson,~S.; Saleh,~O.~A. A
  thin permeable-membrane device for single-molecule manipulation. \emph{Review
  of Scientific Instruments} \textbf{2016}, \emph{87}, 014301\relax
\mciteBstWouldAddEndPuncttrue
\mciteSetBstMidEndSepPunct{\mcitedefaultmidpunct}
{\mcitedefaultendpunct}{\mcitedefaultseppunct}\relax
\EndOfBibitem
\bibitem[Williams(2002)]{williams2002optical}
Williams,~M.~C. Optical tweezers: measuring piconewton forces. \emph{Biophysics
  Textbook Online: http://www. biophysics. org/btol} \textbf{2002}, \relax
\mciteBstWouldAddEndPunctfalse
\mciteSetBstMidEndSepPunct{\mcitedefaultmidpunct}
{}{\mcitedefaultseppunct}\relax
\EndOfBibitem
\bibitem[Schnurr \latin{et~al.}(1997)Schnurr, Gittes, MacKintosh, and
  Schmidt]{schnurr1997determining}
Schnurr,~B.; Gittes,~F.; MacKintosh,~F.; Schmidt,~C. Determining microscopic
  viscoelasticity in flexible and semiflexible polymer networks from thermal
  fluctuations. \emph{Macromolecules} \textbf{1997}, \emph{30},
  7781--7792\relax
\mciteBstWouldAddEndPuncttrue
\mciteSetBstMidEndSepPunct{\mcitedefaultmidpunct}
{\mcitedefaultendpunct}{\mcitedefaultseppunct}\relax
\EndOfBibitem
\bibitem[Wachsstock \latin{et~al.}(1994)Wachsstock, Schwarz, and
  Pollard]{wachsstock1994cross}
Wachsstock,~D.~H.; Schwarz,~W.; Pollard,~T. Cross-linker dynamics determine the
  mechanical properties of actin gels. \emph{Biophysical journal}
  \textbf{1994}, \emph{66}, 801--809\relax
\mciteBstWouldAddEndPuncttrue
\mciteSetBstMidEndSepPunct{\mcitedefaultmidpunct}
{\mcitedefaultendpunct}{\mcitedefaultseppunct}\relax
\EndOfBibitem
\bibitem[Goley and Welch(2006)Goley, and Welch]{nucleationref1}
Goley,~E.~D.; Welch,~M.~D. The ARP2/3 complex: an actin nucleator comes of age.
  \emph{Nat Rev Mol Cell Biol} \textbf{2006}, \emph{7}, 713--726\relax
\mciteBstWouldAddEndPuncttrue
\mciteSetBstMidEndSepPunct{\mcitedefaultmidpunct}
{\mcitedefaultendpunct}{\mcitedefaultseppunct}\relax
\EndOfBibitem
\bibitem[Kaneshiro(2011)]{nucleationref2}
Kaneshiro,~E., Ed. \emph{Cell Physiology Source Book}; Academic Press London ;
  New York, 2011; p 996\relax
\mciteBstWouldAddEndPuncttrue
\mciteSetBstMidEndSepPunct{\mcitedefaultmidpunct}
{\mcitedefaultendpunct}{\mcitedefaultseppunct}\relax
\EndOfBibitem
\bibitem[Burlacu \latin{et~al.}(1992)Burlacu, Janmey, and
  Borejdo]{burlacu1992distribution}
Burlacu,~S.; Janmey,~P.; Borejdo,~J. Distribution of actin filament lengths
  measured by fluorescence microscopy. \emph{American Journal of
  Physiology-Cell Physiology} \textbf{1992}, \emph{262}, C569--C577\relax
\mciteBstWouldAddEndPuncttrue
\mciteSetBstMidEndSepPunct{\mcitedefaultmidpunct}
{\mcitedefaultendpunct}{\mcitedefaultseppunct}\relax
\EndOfBibitem
\bibitem[Gurmessa \latin{et~al.}(2016)Gurmessa, Fitzpatrick, Falzone, and
  Robertson-Anderson]{gurmessa2016entanglement}
Gurmessa,~B.; Fitzpatrick,~R.; Falzone,~T.~T.; Robertson-Anderson,~R.~M.
  Entanglement Density Tunes Microscale Nonlinear Response of Entangled Actin.
  \emph{Macromolecules} \textbf{2016}, \relax
\mciteBstWouldAddEndPunctfalse
\mciteSetBstMidEndSepPunct{\mcitedefaultmidpunct}
{}{\mcitedefaultseppunct}\relax
\EndOfBibitem
\bibitem[Carlier \latin{et~al.}(1987)Carlier, Pantaloni, and
  Korn]{carlier1987mechanisms}
Carlier,~M.; Pantaloni,~D.; Korn,~E. The mechanisms of ATP hydrolysis
  accompanying the polymerization of Mg-actin and Ca-actin. \emph{Journal of
  Biological Chemistry} \textbf{1987}, \emph{262}, 3052--3059\relax
\mciteBstWouldAddEndPuncttrue
\mciteSetBstMidEndSepPunct{\mcitedefaultmidpunct}
{\mcitedefaultendpunct}{\mcitedefaultseppunct}\relax
\EndOfBibitem
\end{mcitethebibliography}

\end{document}